# The role of quantum expansion in cosmic evolution

A D Ernest

*Faculty of Science, room 271, building 30, Locked Bag 588, Charles Sturt University, Wagga Wagga, NSW, 2678, Australia*

*Email: aernest@csu.edu.au*

**Abstract**

A quantum expansion parameter, analogous to the Hubble parameter in cosmology, is defined for a free particle quantum wavefunction. By considering the universe as an initial single Gaussian quantum wavepacket whose mass is that of present-day observable universe and whose size is that of the Planck Length at the Planck Time, it is demonstrated that this quantum expansion parameter has a value at the present epoch of the same order as the value of the Hubble constant. The coincidence suggests examining the effect of including this type of quantum wave expansion in traditional general relativistic cosmology and a sample model illustrating this is presented here. Using standard Einstein-de Sitter cosmology ($\Omega_m = 1$) it is found that cosmic acceleration (aka dark energy) arises naturally during cosmic history. The time at which the universe switched from deceleration to acceleration (observationally ~7 Gyr before the present epoch) yields a value for the mass of the wavepacket representing the universe at the Planck Time and its present age. This same mass may then be used to obtain a curve for the cosmic expansion rate versus *z*. This curve is well fit to observational data. The model is used also to obtain an estimate of the inflationary expansion factor.

## 1. INTRODUCTION

It is now well-established that the universe is accelerating at the present epoch, an observation that cannot be explained solely within the framework of traditional general relativistic cosmology. There is a demonstrable need therefore to consider factors in addition to relativistic space-time that might influence the rate of separation of structures such as galactic clusters. Whilst the cosmological constant enables the general relativistic equations to predict cosmic acceleration, its reintroduction has been somewhat ad hoc and in itself does not offer any satisfactory physical rationale for acceleration. The aim here is to explore an alternative hypothesis whose origins are based on already established physics, in this case quantum theory. Quantum mechanics has been enormously successful in describing the behaviour of nuclei, atoms and molecules as well as a host of larger scale novel phenomena such as quantum connectivity and macroscopically entangled states. Additionally the existence of gravitational quantum eigenstates has been experimentally demonstrated (Nesvizhevsky et al. 2002), and theoretical considerations (Ernest 2009 a, b, 2012) suggest that such states may be connected with dark matter. These successes of quantum theory, involving different types of potentials over a range of scales, suggest that it might also be implicated in other cosmological phenomena such as the evolution of the universe. This paper draws on a predicted property of quantum wavefunctions that, like the cosmological constant, can be also shown to lead cosmic acceleration when incorporated into a general relativistic model.

It is assumed here that the net recession of galactic clusters originates from two separate phenomena: (1) that due to traditional relativistic cosmology in flat-space (as per WMAP observations (Larson et al. 2011) and (2) an additional component resulting from the prediction that quantum theory makes about the behaviour of the universe if treated as a single quantum wavefunction. In this latter assumption the wavefunction of the universe is pictured as having a mass which begins at the Planck Time ($5.4 \times 10^{-44}$ s) as a very small dense clump of matter, and is represented by a single compact wavefunction, the width of which is the Planck Length ($1.6 \times 10^{-35}$ m), the Planck Length and Time being natural choices for the initial conditions of the wavefunction. It is then possible to work out how this wavefunction should evolve over time according to conventional quantum physics and incorporate its development into cosmic evolution. This is accomplished by defining a time-dependent expression for a 'specific quantum expansion rate' of the wavefunction, analogous to the Hubble parameter in the general relativistic approach. The Hubble terms are used to derive infinitesimal changes in the separation of free objects in the universe due to each component and added on the basis that the quantum function representing the universe is a freely expanding wave superimposed on a space that is already expanding.

The background theory is given in section 2. Section 3 defines the concept of the specific quantum expansion rate and it is shown that, for an initial wavefunction whose mass is taken as the order of that of the observable universe, the present day specific quantum expansion rate is remarkably similar to the present day value of the Hubble constant. A mass of this order might therefore be an appropriate initial value to use in any model that might predict cosmic acceleration. Section 4 presents a 'proof of principle' that such a hypothesis can lead to cosmic acceleration by developing a simple example of hybrid quantum-relativistic cosmology that includes quantum wavepacket expansion (and Einstein-de Sitter cosmology in this case). The time dependence of the cosmological deceleration parameter $q$ is calculated in this hybrid quantum-Einstein-de Sitter model, and it will be shown that there is a relation between the mass of the initial wavefunction, the present age of the universe and the time at which the universe switched from a decelerating to accelerating phase. The evolution of the cosmic expansion rate is also calculated and compared with that derived from data based on supernovae observations.

## 2. BACKGROUND

In quantum theory a free particle wavefunction expands at a rate that depends on its initial mass and size. The mathematical treatment of this is covered in most standard texts (Schiff, 1968) and the approach is summarised here. The initial wavefunction $\psi(x,y,z,t)$ of the universe will be represented by a minimal-uncertainty Gaussian wavepacket which is the most easy to handle mathematically. (This potentially introduces interpretational difficulties and these will be discussed later.) The form of such a packet at $t=0$ is described separately in

each of the three orthogonal directions by a localised (and normalised) function

$\psi(x,0) = (1/2\pi\Delta x_0^2)^{1/4} \exp[-(x-\bar{x})^2/4\Delta x_0^2 + i\bar{k}x]$, where $\bar{x}$ is the average packet position, $\bar{k}$ the average packet wavenumber and $\Delta x_0$ the width parameter at $t=0$ of the corresponding probability density function, $\psi^*\psi = (1/2\pi^2\Delta x_0^2)^{1/4} \exp\left[-(x-\bar{x})^2/2\Delta x_0^2\right]$. $\psi(x,0)$ may be decomposed into a weighted spectral distribution function $\varphi(k,t)$ of the field-free time dependent momentum eigenfunctions

$u_k(x,t) = (1/2\pi)^{1/2} \exp(ikx - i\hbar k^2 t/2m) = u_k(x)\exp(-i\hbar k^2 t/2m)$ in $k$-space (wavenumber-space) at $t=0$ using $\varphi(k,0) = \int_{-\infty}^{\infty} (1/2\pi^2\Delta x_0^2)^{1/4} \exp[-(x-\bar{x})^2/4\Delta x_0^2 + i(\bar{k}-k)x]dx$. The temporal evolution of $\psi(x,t)$ is then obtained using $\psi(x,t) = \int_{-\infty}^{\infty} \varphi(k,0)u_k(x)\exp(-i\hbar k^2 t/2m)dk$ to give

$$\psi(x,t) = \left(2\Delta x_0 + \frac{i\hbar t}{m\Delta x_0}\right)^{-1/2} \left(\frac{2}{\pi}\right)^{\frac{1}{4}} \exp\left[-\frac{m\left((x-\bar{x})^2 - 4i\bar{k}x\Delta x_0^2\right) + 2\bar{k}\left(\bar{x} + i\bar{k}\Delta x_0^2\right)\hbar t}{4m\Delta x_0^2 + 2i\hbar t}\right], \quad (1)$$

and probability density

$$\psi^*(x,t)\psi(x,t) = \left(4\Delta x_0^2 + \frac{\hbar^2 t^2}{m^2\Delta x_0^2}\right)^{-1/2} \left(\frac{2}{\pi}\right)^{\frac{1}{2}} \exp\left[-\frac{2\Delta x_0^2\left(m(\bar{x}-x) + \bar{k}\hbar t\right)^2}{4m^2\Delta x_0^4 + \hbar^2 t^2}\right]. \quad (2)$$

From equations (1) and (2) it is clear that the wavepacket retains its Gaussian form with time although it is no longer minimal (that is the position-momentum uncertainty product increases with time), because the width of the corresponding spectral distributions $\varphi(k,t)$ and $\varphi^*(k,t)\varphi(k,t)$ are independent of time. From equation (2) the width parameter (generally symbolised by a standard deviation $\sigma$ but here by $\Delta x(t)$) of the probability density of the wavepacket varies with time as

$$\Delta x(t) = \left( \Delta x_0^2 + \frac{\hbar^2 t^2}{4m^2 \Delta x_0^2} \right)^{1/2} = \Delta x_0 \left( 1 + \frac{\hbar^2 t^2}{4m^2 \Delta x_0^4} \right)^{1/2} = \Delta x_0 \left( 1 + \frac{t^2}{\tau^2} \right)^{1/2}, \quad (3)$$

where $\tau$, given by

$$\tau = 2m\Delta x_0^2 / \hbar \quad (4)$$

can be considered a characteristic time relating to the initial conditions of the wavefunction. An initial wavefunction with $\sigma = \Delta x_0$ expands at a gradually increasing rate until, for times $t \gg \tau$, the rate approaches the constant value of $\Delta x_0/\tau$. The rate is larger the smaller the mass of the entity and the more localised the initial packet conditions.

The value of $\tau$ for wavepackets representing elementary particles is generally relatively short so that a significant change in the size of their wavefunctions occurs over any realistically 'macroscopic' time, reflecting the increase in position uncertainty prior to a subsequent, usually localising, position measurement. Simple, small-mass, 'compound' objects such as atoms behave in a similar way but their internal wavefunction is bound however, so no quantum expansion takes place internally. (Indeed bound internal quantum wavefunctions such as those describing the internal structure of objects such as atoms, or the binding in galaxies, are resilient even to relativistic expansion since space expands through these structures.) Conversely, for classical macroscopic objects, $\tau$ is usually very long and the corresponding expansion rate extremely small, undetectable over the lifetime of the universe: $m = 1$ kg and $\sigma = \Delta x_0 = 0.1$ m implies $\Delta x_0/\tau$ ~5 x $10^{-34}$ m/s. In the present work the universe is represented as a single Gaussian packet and it can be shown by suitably modifying the Gaussian profile, that if such a packet has a perturbed shape consisting of an array of localised 'peaks' of probability density superimposed on the Gaussian function (such as that which might describe the universe after internally it had developed to form galactic clusters) then, as time proceeds, the average position of these peaks separate from each other as part of the ongoing spread of the overall packet, in a similar way to the relativistic expansion of the universe.

## 3. SPECIFIC EXPANSION RATE AND THE UNIVERSAL WAVEFUNCTION

As described in section 1, is useful to define a specific quantum expansion rate $H_Q$ for $\Delta x(t)$ given by $H_Q = \frac{d\Delta x(t)}{dt}(\Delta x(t))^{-1}$. Using equations (3) and (4) enables $H_Q$ to be written as

$$H_Q(t,\tau) = \frac{d\Delta x(t)}{dt}(\Delta x(t))^{-1} = \frac{t}{\tau^2 + t^2} \tag{5}$$

$H_Q$ evolves linearly as $t/\tau^2$ for $t \ll \tau$ and as $1/t$ for $t \gg \tau$, that is independently of the initial mass or size of the interaction.

One might expect from the discussion in section 2 that the expansion of free particle quantum wavefunctions could have no relevance for macroscopic objects and certainly no relevance to large scale cosmology or the expansion of the universe. However given that $H_Q$ is essentially a Hubble parameter for quantum wavepackets, it is interesting to compare the present day values of these two quantities when the universe is considered as a single initial quantum wavefunction. The initial volume of this wavefunction is taken as determined by the Planck length $\ell_P = (\hbar G/c^3)^{1/2}$, leading to an initial width parameter $\Delta x_0$ given by $\Delta x_0 = \ell_P/2 = (\hbar G/c^3)^{1/2}/2$. The value of $H_Q$ also depends also on the initial wavefunction mass $m$ through $\tau$. A natural initial trial value for the mass is that of the observable universe, calculated using an estimate of its observable volume of 3 x $10^{80}$ m$^3$ (based on flat space geometry (Larson et al. 2011) and the critical density 9.3 x $10^{-27}$ kg m$^3$ (Mo et al. 2010) Depending on whether $m$ is taken as the total mass including dark energy, the matter content or just the baryonic component, the $m$ values are 2.8 x $10^{54}$, 8.4 x $10^{53}$, 1.7 x $10^{53}$ kg (using the WMAP results (Larson et al. 2011), and values of $H_Q$ as 3.4 x $10^{-20}$, 3.3 x $10^{-19}$, and 1.9 x $10^{-18}$ $s^{-1}$ respectively (taking the present epoch as $t = t_0 \sim 4\times10^{17} s$). The observed Hubble constant is $H \equiv H_R = 71$ km $s^{-1}$ $Mpc^{-1}$ or $2.3\times10^{-18}$ $s^{-1}$. The similarities in the order of magnitude of $H_Q$ to that of $H_R$ are intriguing since they predict the existence of a Hubble flow due to quantum effects from the earliest times that is of the same order at the present epoch as that due to general relativity. Furthermore it means that quantum effects originating from representing the universe as a single primordial wavepacket could significantly affect the rate of expansion in the universe's recent history. Curves showing the variation of $H_Q$ over cosmic history for various masses are given in Figure 1. Of course no allowance for a 'quantum flow' is taken into account in the Friedmann equations and it would seem important therefore to develop and investigate the properties of cosmological models that include this effect, particularly given the similarity between the quantum and relativistic Hubble parameters in the present epoch.

## 4. INCLUSION OF QUANTUM EFFECTS IN COSMOLOGICAL MODELS

The approach adopted here assumes a standard Einstein-de Sitter cosmology in a matter dominated universe given by the Friedmann equation (Liddle, 2003) as

$$\left(H(t)\right)^2 = \left(\frac{\dot{a}(t)}{a(t)}\right)^2 = \frac{8\pi G}{3}\rho(t) - \frac{k}{a(t)^2} \qquad (6)$$

where $H(t)$ is Hubble's constant, $a(t)$ the scale factor corresponding the relative expansion of the universe with time, $\rho(t)$ the density and $k$ the curvature parameter equal to zero in the cosmology being adopted here. Using the equation of state

$$\dot{\rho}(t) + 3\frac{\dot{a}(t)}{a(t)}\left(\rho(t) + \frac{p(t)}{c^2}\right) = 0 \tag{7}$$

and taking the pressure term $p(t) = 0$ in the present matter dominated era, gives the standard result for the temporal variation of the Hubble constant in this exclusively general relativistic model as

$$H(t) = H_{GR} = \frac{2}{3t} \tag{8}$$

The quantum and general relativistic Hubble parameters given in equations (5) and (8) respectively each describe the infinitesimal change in the relative separation over time $dt$ that occurs between any two 'unbound' objects in the universe whose separation $x(t)$. The combined infinitesimal increase $dx$ in separation for both processes is therefore given by that due to the addition of both processes:

(1) the quantum contribution described by $H_Q(t,\tau)$ in equation (5), with now $x(t) \equiv \Delta x(t)$ corresponding to *any* arbitrary separation within the Gaussian profile, plus

(2) the contribution due to the standard relativistic term. The total infinitesimal displacement $dx$ is thus given by

$$dx = dx_Q + dx_{GR} = v_Q dt + v_{GR} dt = \left(H_Q(t) + H_{GR}(t)\right)x(t)dt \tag{9}$$

where $v_Q$ and $v_{GR}$ are effective quantum and relativistic expansion 'velocities'. This leads to

$$H(t) = \left(\frac{dx(t)}{dt}\right)_{total} \bigg/ x(t) = H_Q(t) + H_{GR}(t), \tag{10}$$

where $H(t)$ is the Hubble term for both processes. Using equations (5), (6) and (8) gives

$$\frac{\dot{a}(t)}{a(t)} = H(t) = H_{GR} + H_Q = \frac{2}{3t} + \frac{t}{\tau^2 + t^2}. \tag{11}$$

The cosmological deceleration parameter $q$ is given by $q = -\ddot{a}/aH^2$. Differentiating equation (11) gives the acceleration $\ddot{a}(t)$ as

$$\frac{\ddot{a}}{a} = \left(\frac{2}{3t} + \frac{t}{\tau^2 + t^2}\right)^2 + \left(-\frac{2}{3t^2} + \frac{\tau^2 - t^2}{\tau^2 + t^2}\right), \tag{12}$$

and hence $q$ may be written as

$$q = -\frac{\ddot{a}}{a}\frac{1}{H^2} = -\frac{10t^4 + 17t^2\tau^2 - 2\tau^4}{\left(5t^2 + 2\tau^2\right)^2}. \tag{13}$$

The present day value of the Hubble constant $H(t) = H(t_0) = 2.30 \times 10^{-18}$ s$^{-1}$, accurate to approximately 3% (Larson et al. 2011), fixes a relation between the value of $\tau$ and the present age of the universe $t_0$ and $H_0$ via equation (11) as

$$H(t_0) = H_0 = \frac{2}{3t_0} + \frac{t_0}{\tau^2 + t_0^2} \quad \text{so that} \quad \tau = t_0 \left(-\frac{H_0 t_0 - 5/3}{H_0 t_0 - 2/3}\right)^{1/2}. \tag{14}$$

It is useful to write $t_0$ in equation (14) in terms of the other quantities so that the age of the universe may be found from the value of $\tau$. It is clear that real solutions to equation (14) are only possible when $5/3 > H_0 t_0 > 2/3$. This fixes the value of $t_0$ in terms of $\tau$ and the corresponding initial mass. Given that the value of $H_0$ is robust, the range allowable universal ages is then between $2.9 \times 10^{17}$ and $7.2 \times 10^{17}$ s, and the range of masses corresponding to these ages extends from $3.2 \times 10^{54}$ down to $1.5 \times 10^{52}$ kg. By substituting $\tau = 2m\Delta x_0^2/\hbar = m\ell_P^2/(2\hbar)$ into equation (14) and solving, $t_0$ may be alternatively expressed in terms of the initial mass $m$ and the Planck Length $\ell_P$ as

$$t_0 = \frac{5}{9H_0}$$

$$+ \frac{100\hbar^2 - 27H_0^2 l_P^4 m^2 + \left(1000\hbar^3 + 81H_0^2 l_P^4 m^2 \hbar + 9\sqrt{3}H_0 l_P^2 m\sqrt{81H_0^4 l_P^8 m^4 - 873H_0^2 l_P^4 m^2 \hbar^2 + 4000\hbar^4}\right)^{2/3}}{18H_0 \left(1000\hbar^6 + 81H_0^2 l_P^4 m^2 \hbar^4 + 9\sqrt{3}H_0 l_P^2 m\hbar^3 \sqrt{81H_0^4 l_P^8 m^4 - 873H_0^2 l_P^4 m^2 \hbar^2 + 4000\hbar^4}\right)^{1/3}}$$

(15)

Equation (15) comes directly from the solution to the cubic equation derived from the rearrangement of equation (14). Figure 2 shows the value of the deceleration parameter $q$ as a function of time for the range of allowable initial masses consistent with the presently accepted value of the Hubble constant. Curve (a) corresponds to a sufficiently large wavefunction mass ($6.5 \times 10^{55}$ kg shown here) that quantum expansion has no effect on the original Einstein-de Sitter space, $q$ remains constant at 0.5 and $t_0 = 2/3H_0$. As the initial wavefunction mass is decreased, the values of $q$ begin to decrease with time due to the effect of quantum expansion. This results in curves like those of curve set (b) in Figure 2. These curves, when interpreted purely by general relativity without quantum effects, mimic universes that have mass density parameters $\Omega_m$ less than the closure density and require the introduction of dark energy or a cosmological constant. With quantum expansion however negative $q$ values arise as a natural consequence of cosmic evolution and a cosmological constant is no longer required. A mass of $\sim 8.0 \times 10^{53}$ kg is required to produce a curve in this model that results in acceleration beginning to be observed at the present epoch. Further reduction in initial mass results in a universe that spends a progressively longer relative fraction of its history in an accelerated mode.

Using the fact that infinitesimal rate of expansion of a photon at any time $t$ may be related to the value of $H(t)$ via the equation $d\lambda(t)/\lambda(t) = H(t)dt$ and knowing the functional form of $H(t)$ from equation (11), expressions for the scale factor $a(t)/a(t_0)$ and redshift $z$ may be derived for the hybrid quantum-relativistic model presented above to give

$$\frac{a(t)}{a(t_0)} = \left(\frac{t}{t_0}\right)^{2/3} \left(\frac{t^2 + \tau^2}{t_0^2 + \tau^2}\right)^{1/2} \text{ and } z = \frac{\lambda_e}{\lambda_0} - 1 = \left(\frac{t_e}{t_0}\right)^{2/3} \left(\frac{t_e^2 + \tau^2}{t_0^2 + \tau^2}\right)^{1/2} - 1 \qquad (16)$$

where $\lambda_0$ is the wavelength of the received photon at the present time $t_0$, $\lambda_e$ is the wavelength at the time of emission $t_e$. Any cosmic time $t_e$ may be related to a value of $z$ and vice versa using equation (16). The results from observations of type 1a supernova suggest that the acceleration of the universe began at a redshift of $z \sim 0.86$ with 95% confidence (Ishida et al. 2007, Amanullah et al. 2010). Using equation (16) this value of $z$ corresponds to the time of transition from a decelerated to accelerated phase of the universe ($q = 0$) as occurring at approximately 7 Gyr before the present epoch. The curve of Figure 2 corresponding to this time is that labelled * and corresponds to an initial wavepacket mass of $5 \times 10^{53}$ kg and gives the present age of the universe as $4.3 \times 10^{17}$ s.

Figure 3 shows the predicted $\dot{a}(z)$ versus $z$ curve for the present quantum-Einstein- de Sitter model using the *same* mass value that was used in figure 2 to obtain the observed switch-over redshift of $z \sim 0.86$ (Ishida et al. 2007 and Amanullah et al. 2010). The figure also shows predicted curves for other theoretical models, in particular a standard Einstein-de Sitter model with $\Omega_m = 1$, and the $\Lambda$CDM model with a dark energy component and $\Omega_m = 0.27$. Also shown are observationally derived values of $\dot{a}(z)$ obtained by incorporating the supernovae/WiggleZ project results (Blake et al. 2011). The present quantum-Einstein- de Sitter model is well fit to these data, as is that predicted by $\Lambda$CDM cosmology. Significantly in the quantum model however, dark energy arises as a natural evolution of quantum wavepacket expansion rather than the use of an artificial and essentially ad hoc addition of a cosmological constant as in the traditional dark energy model. The universe remains flat using Einstein-de Sitter space-time and $\Omega_m = 1$. Since both the cosmological constant model and the quantum-Einstein-de Sitter model agree well with the work of Blake et al. it is not possible to distinguish between these two models with the present data.

The mass of $5\times10^{53}$ kg represents the wavepacket mass at the Planck Time and also the approximate mass of the observable universe at the present epoch. Hence the size of the observable universe and the size of the Planck Length represent the expansion factor that has occurred over the time $t=5.4\times10^{-44}$ to $t=4\times10^{17}$ s. Equation (16) quantifies the expansion in the quantum-Einstein-de Sitter model, and can be used to calculate the size of the observable universe at any time over which it is valid before the present era, in particular its size at the end of the Grand Unified Theory (GUT) era ($10^{-34}$ s, (Liddle 2003)). Using 8 x $10^{26}$ m as the present size of the wavefunction equation (16) gives the size at $10^{-34}$ s as about 1.7 m, implying that the universe would have needed to expand from the Planck Length at the Planck Time to 1.7 m at ~$10^{-34}$ s, resulting in an expansion factor over this time ~$10^{35}$. Assuming the inflationary scale factor follows $a(t)=\exp\left((\Lambda/3)^{1/2}t\right)$ (Liddle 2003), and assuming the GUT era lasted a substantial fraction of the first $10^{-34}$ s, the estimated value of $\Lambda$ is 2 x $10^4$ s$^{-2}$.

Two improvements on the present treatment would be (1) to adopt a profile so that the mass of the universe is fully contained within the initial Planck Length and (2) use a profile which is better representative of a uniform density isotropic universe than a Gaussian. The Gaussian profile is self-similar under quantum expansion and its use considerably simplifies the mathematics. It is expected that a model which incorporated features (1) and (2) would yield a best –fit initial wavefunction mass marginally larger than that incorporating the Gaussian function but further work is required to examine the degree to which the parameter values would change. Investigations examining this are continuing.

The Planck Length and Planck Time are assumed to represent the smallest units of space and time that have physical meaning. In the present model only one spatial Planck volume was examined but there was no a-priori need that the entire universe consists solely of one Planck unit of space, and nothing to prevent the entire universe being composed of a (possibly infinite) number of adjacent units of space occupied by similar masses to give an isotropic, homogeneous and possibly infinite universe. In this scenario each quantum unit would behave in a similar way. During the inflationary period and early universal times, relativistic expansion dominates and these units would remain adjacent but isolated from each other. More recently in cosmic history however, within the last several billion years, quantum expansion dominates and there would be potential for these adjacent regions to merge. Such regions could be evidenced by anisotropies in the observed Hubble flow at large cosmic distances (Kashlinsky 2008).

## 5. CONCLUSION

In this paper traditional quantum theory has been used to predict a rate for the quantum expansion of the universe if it was represented as an initial single wavepacket beginning at the Planck Time and having a size equal to the Planck Length. The predicted rate obtained is of the same order as the Hubble constant if a mass of the order of that of the observed universe is used for the initial wavefunction. It was shown using these initial conditions, that quantum theory predicts a significant component to universal expansion which is not included in the existing general relativistic approach when the universe is treated in this way. The evidence for the existence of an initial universal wavefunction this is made more compelling by the fact that when the quantum expansion term is included in flat-space cosmology the resulting hybrid model predicts the onset of cosmic acceleration without the need for an ad hoc cosmological constant and dark energy as such. In the simple flat-space Einstein-de Sitter universe with $\Omega_m \sim 1$ modelled here, it was possible to obtain an expected value of the deceleration-acceleration transition point at ~7 Gyr before present time using a mass of the observable universe of $5\times10^{53}$ kg. A value for the expansion factor during the inflationary period was found to be ~$10^{35}$ and the parameter $\Lambda$ estimated at $2 \times 10^4$ s$^{-2}$.

Acknowledgements

The author would like to thank Prof Neville Fletcher for his valuable suggestions and comments, and Dr Hans Swan and Dr Matthew Collins for their proof reading of the manuscript.

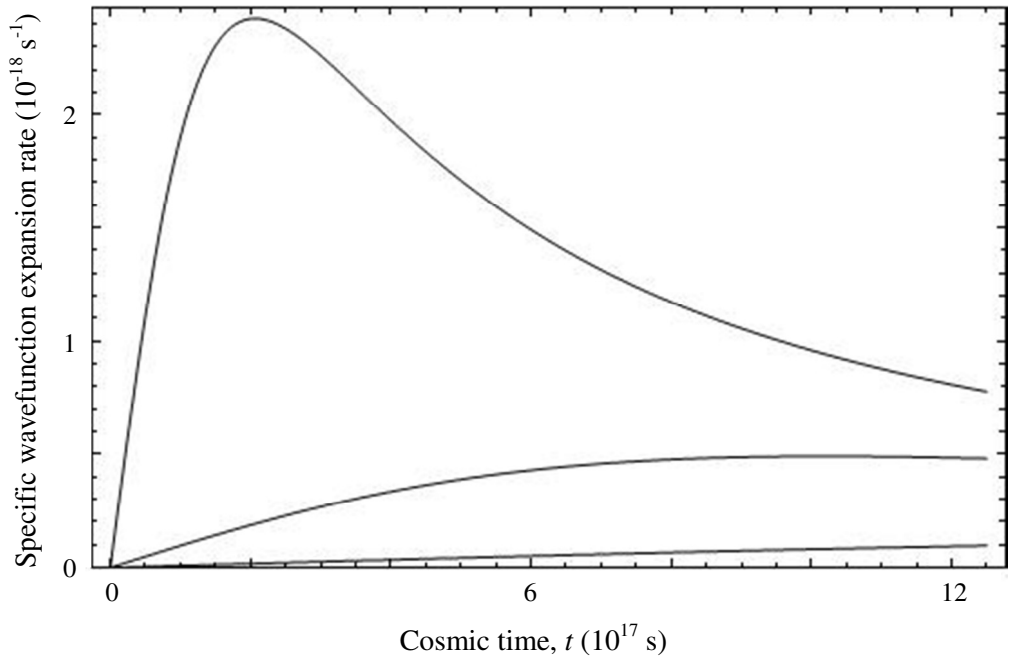

Fig. 1 Variation of the specific expansion rate, equivalent to a Hubble-type parameter, for various values of $\tau$. Each wavefunction originates at $t \sim 0$ with width $2\Delta x_0$ equal to the Planck length, $1.616 \times 10^{-35}$ m. Top, middle and lower curves are $\tau = 3.4$, $1.0$ and $0.21$ ($\times 10^{18}$) s respectively, corresponding to initial mass values of 2.8, 0.84 and 0.17 ($\times 10^{54}$) kg.

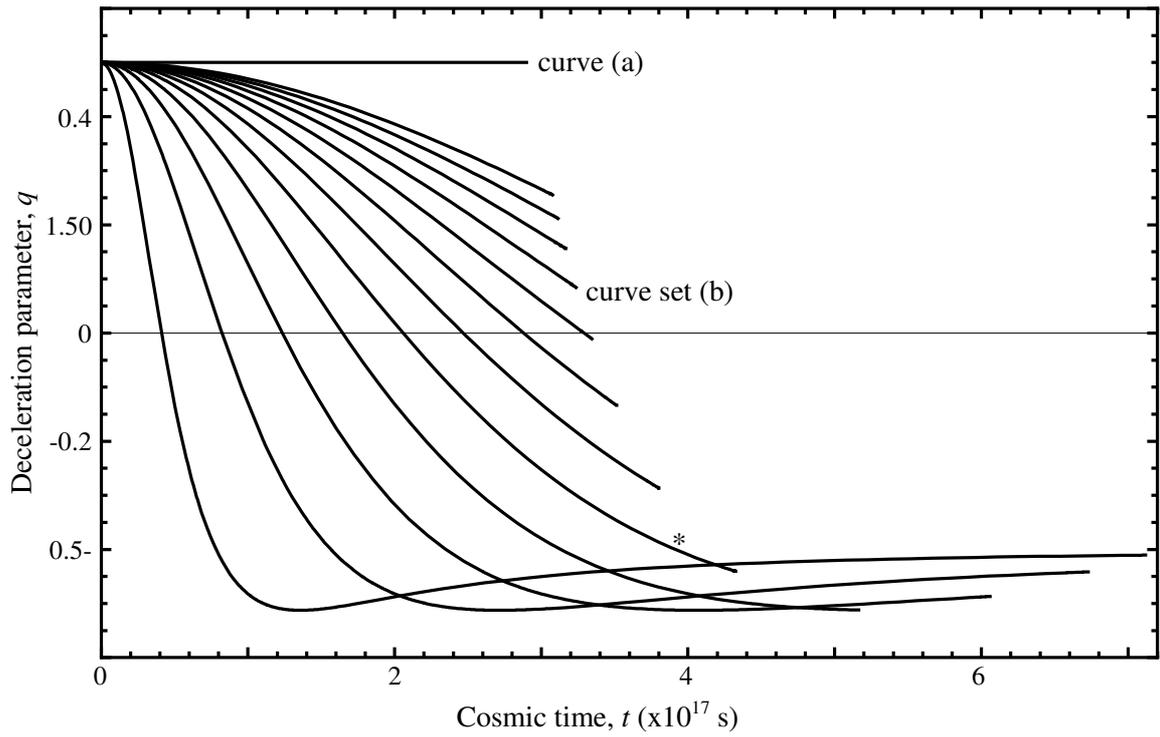

Fig. 2. Plots of the deceleration parameter $q$ as a function of cosmic time for the quantum-Einstein-de Sitter universe ($\Omega_m = 1$). Each curve corresponds to a wavepacket of given initial mass consistent with the present value of the Hubble constant. Each curve finishes at the present era, a time which varies depending on the initial mass.

Curve (a): $6.5 \times 10^{55}$ kg; curve set (b): first (upper) curve $1.2 \times 10^{54}$ kg, then each successive curve with mass decreasing by $1.0 \times 10^{53}$ kg. The starred curve (*) is the best fit curve to $q = 0$ at $z = 0.86$.

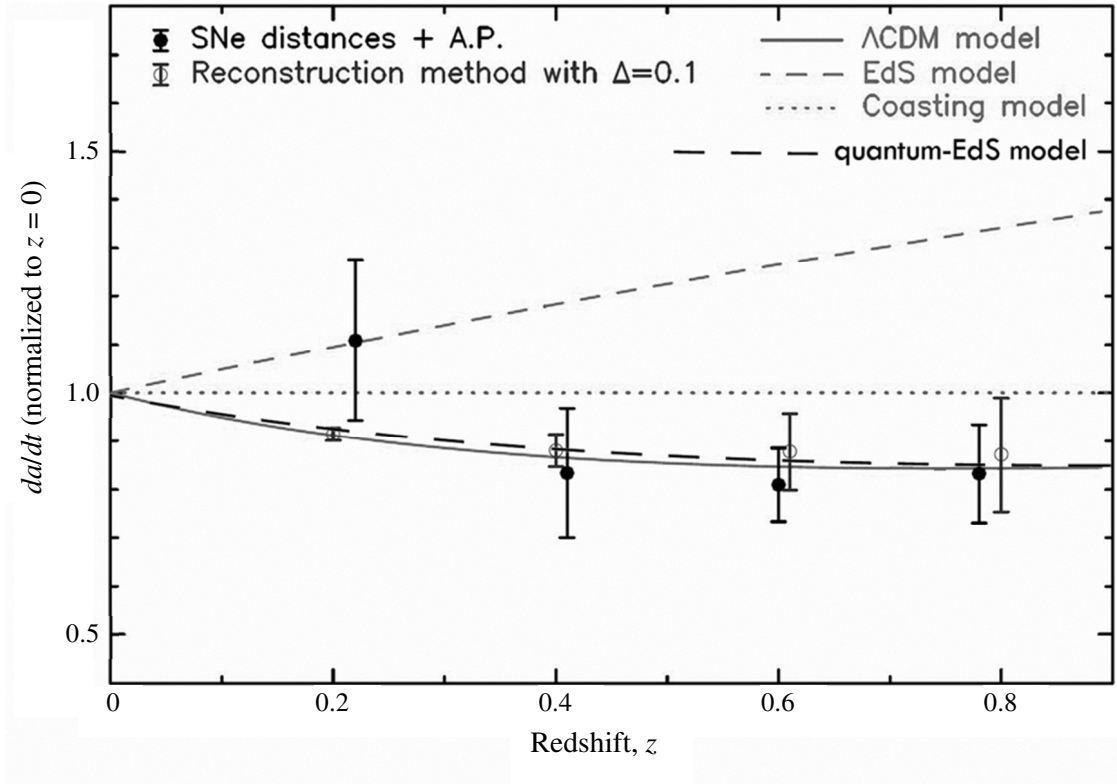

Fig 3. Values of $\dot{a}(t)(\equiv da/dt)$ for the quantum-Einstein-de Sitter model superimposed on the measurement of the evolution of the cosmic expansion rate using Alcock-Paczynski and supernovae data (solid and open points, Blake et al. 2011, figure reproduced with permission, Copyright Clearance Centre and MNRAS). The quantum-Einstein-de Sitter mass parameter is $5\times10^{53}$ kg, the same mass as that corresponding to the deceleration-acceleration switch over time at $z = 0.86$ of figure 2. The figure shows that inclusion of quantum wavepacket expansion in a flat-space Einstein-de Sitter universe ($\Omega_m = 1$) can successfully predict the results of supernovae observations, and mimic $\Lambda$CDM cosmology with a dark energy component and $\Omega_m = 0.27$.